\documentclass[prl,showpacs,floatfix,twocolumn,superscriptaddress]{revtex4}
\usepackage{graphicx}
\usepackage{dcolumn}

\begin{document}

\newcommand*{\cm}{cm$^{-1}$\,}
\newcommand*{\tise}{1$T$-TiSe$_2$\,}

%
\title{Anomalous metallic state of Cu$_{0.07}$TiSe$_2$: an optical spectroscopy study}
%
%
\author{G. Li}
\author{W. Z. Hu}
\author{J. Dong}
\affiliation{Beijing National Laboratory for Condensed Matter
Physics, Institute of Physics, Chinese Academy of Sciences,
Beijing 100080, P. R. China}
\author{D. Qian}
\author{D. Hsieh}
\author{M. Z. Hasan}
\affiliation{Department of Physics, Joseph Henry Laboratories of
Physics, Princeton University, Princeton, New Jersey 08544, USA}

\author{E. Morosan}
\author{R. J. Cava}
\affiliation{Department of Chemistry, Princeton University,
Princeton, New Jersey 08544, USA}

\author{N. L. Wang}
\email{nlwang@aphy.iphy.ac.cn}%
\affiliation{Beijing National Laboratory for Condensed Matter
Physics, Institute of Physics, Chinese Academy of Sciences,
Beijing 100080, P. R. China}
%
%
%

\begin{abstract}
We report an optical spectroscopy study on the newly discovered
superconductor Cu$_{0.07}$TiSe$_2$. Consistent with the
development from a semimetal or semiconductor with a very small
indirect energy gap upon doping TiSe$_2$, it is found that the
compound has a low carrier density. Most remarkably, the study
reveals a substantial shift of the "screened" plasma edge in
reflectance towards high energy with decreasing temperature. This
phenomenon, rarely seen in metals, indicates either a sizeable
increase of the conducting carrier concentration or/and a decrease
of the effective mass of carriers with reducing temperature. We
attribute the shift primarily to the later effect.
\end{abstract}


\maketitle

%

Charge density waves (CDW) and superconductivity are two important
broken symmetry states in solids. The interplay between the two
states has been a topic of central interest in condensed matter
physics. Among various CDW materials, layered \tise is
particularly interesting. The compound shows a lattice instability
around 200 K, below which it enters into a commensurate CDW phase
associated with a ($2\times2\times2$)
superlattice\cite{Wilson1,Wilson2,Salvo}. Unlike the case of most
CDW materials, the CDW transition in this compound is not driven
by Fermi surface nesting. The ground state is believed to be
either a semimetal or a semiconductor with a very small indirect
gap\cite{Bachrach,Traum,Anderson,Pillo,Kidd,Rossnagel,Cui,Li}.
It has recently been reported that, upon controlled
intercalation of Cu into \tise to yield Cu$_x$TiSe$_2$, the CDW
transition is continuously suppressed, and a superconducting state
emerges near x=0.04, with a maximum T$_c$ of 4.15 at
x=0.08\cite{Morosan}. This is the first superconducting system
realized in the 1T structure CDW-bearing family. The CDW-superconductivity phase diagram developed for Cu$_x$TiSe$_2$ is analogous to the
antiferromagnetism-superconductivity phase diagram found for the
high-temperature superconductors. The system offers a good
opportunity to study the evolution of competing electronic states
from CDW to superconductivity.

Recently, high quality single crystals have been successfully
grown for the system. A detailed study of the superconducting
properties has been performed on Cu$_{0.07}$TiSe$_2$ single
crystals\cite{Morosan2}. It is of interest to
investigate the electronic properties of the doped metallic phase.
In this letter, we report the first optical study on the
superconducting compound Cu$_{0.07}$TiSe$_2$. Consistent with the
development from a semimetal or semiconductor with a very small
indirect energy gap upon doping, the study reveals that the
compound has a low carrier density. Most remarkably, a substantial
shift of the plasma edge in reflectance towards high energy with
decreasing temperature is observed. This phenomenon, rarely seen
in metals, indicates either a sizeable increase of the conducting
carrier concentration or/and a decrease of the effective mass of
carriers with reducing T. We conclude that the shift is
mainly due to the later effect.

Superconducting single crystals of Cu$_{0.07}$TiSe$_2$
(T$_c\sim$4K) with optically flat surfaces were grown via chlorine
vapor transport, as described previously\cite{Morosan2}.
The frequency-dependent reflectance spectra R($\omega$) at
different T were measured by a Bruker IFS 66 v/s spectrometer in
the range from 50 to 25,000 cm$^{-1}$ and a grating-type
spectrometer from 20,000 to 50,000 \cm. The sample was mounted on
an optically black cone on the cold-finger of a He flow cryostat.
An \emph{in situ} gold (50 $\sim$ 15,000 cm$^{-1}$) and aluminum
(9,000 $\sim$ 50,000 cm$^{-1}$) overcoating technique was employed
for reflectance measurements. The Kramers-Kronig transformation of
R($\omega$) was used to obtain the other optical response
functions. A Hagen-Rubens relation was used for low frequency
extrapolation, and a constant extrapolation for high-frequency to
300,000 cm$^{-1}$ followed by a function of $\omega^{-2}$ was used
for the higher energy side.

Fig. 1 shows the reflectance spectra at different temperatures:
(a) R($\omega$) spectra over very broad energy scale from 50 to
50,000 \cm on a logarithmic scale; (b) the spectra in an expanded
plot over the frequency range from 50 \cm to 5,000 \cm. The
R($\omega$) values at low $\omega$ are rather high and increase
further with decreasing T. This is a typical metallic response.
With increasing $\omega$, R($\omega$) drops quickly to a minimum
value near 3,000 \cm, usually referred to as the "screened" plasma
edge.\cite{Li} The relatively low edge position reveals a low
carrier density. Most surprisingly, the plasma edge was found to
display a substantial shift towards higher frequencies
(\textit{i.e.} a blueshift) with decreasing T. Above the edge
frequency, the reflectance becomes roughly T-independent. Near
15,000 \cm, there is another edge structure in R($\omega$), which
does not originate from the free-carrier plasmon, but is caused by
an interband transition. Besides these features, a phonon
structure near 140 \cm can be clearly observed at high T. The
feature weakens at low T due to increased screening by free
carriers.

\begin{figure}[t]
\centerline{\includegraphics[width=2.8in]{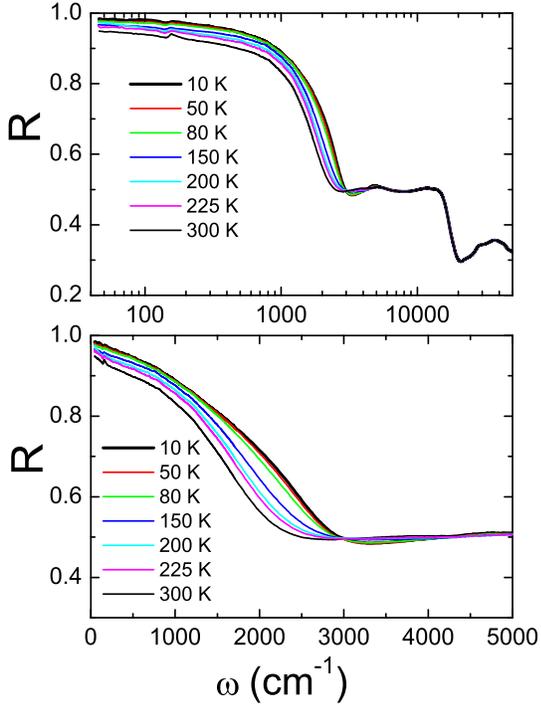}}%
\vspace*{-0.20cm}%
\caption{(Color online) (a) The T-dependent R($\omega$) in the
frequency range from 50 \cm to 50,000 \cm on a logarithmic scale.
(b) The expanded plot of R($\omega$) from 50 \cm to 5,000
\cm.}%
\label{fig1}
\end{figure}

\begin{figure}[b]
\centerline{\includegraphics[width=2.9in]{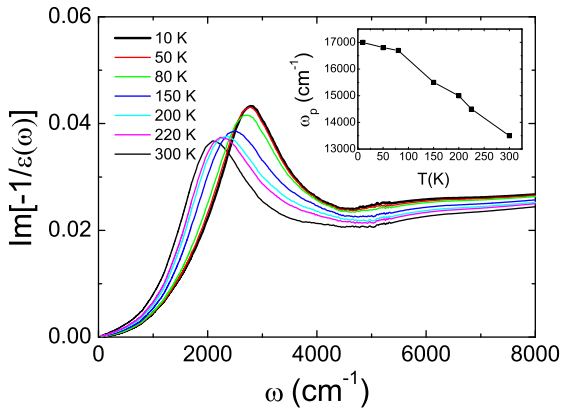}}%
\vspace*{-0.20cm}%
\caption{(Color online) The energy loss function spectra at
different T. Inset: the unscreened plasma
frequency obtained from the fit by equation (1) to reflectance data.}%
\label{fig1}
\end{figure}

The formation of the plasma edge and its evolution with T is also
reflected in the energy loss function,
$Im{[-1/\epsilon(\omega)]}$, as shown in Fig. 2. In the energy
loss function, the screened plasma frequency
($\omega_p'=\omega_p/\sqrt{\epsilon_\infty}$) corresponds to the
peak position, while the carrier damping (or scattering rate) is
linked to the peak width. We can see clearly that, as T decreases
from 300 K to 10 K, the major change is the blueshift of the
screened plasma frequency.

\begin{figure}[t]
\centerline{\includegraphics[width=2.9in]{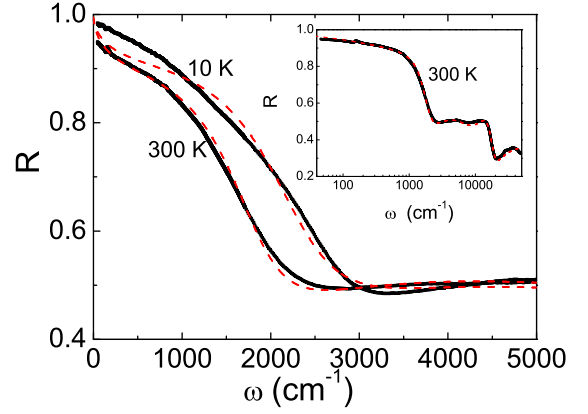}}%
\vspace*{-0.20cm}%
\caption{(Color online) The optical reflectance at 300 K and 10 K,
together with the fit by equation (1).
Inset shows the spectra over a broad frequency range.}%
\label{fig3}
\end{figure}

To analyze the variation of the carrier density and its damping
quantitatively, we fit the experimental reflectance to a simple
Drude-Lorentz model:\cite{Degiorgi2}
\begin{equation}
\epsilon(\omega)=\epsilon_\infty-{{\omega_p^2}\over{\omega^2+i\omega/\tau}}+\sum_{i=1}^2{{S_i^2}\over{\omega_i^2-\omega^2-i\omega/\tau_i}}.
\label{chik}
\end{equation}
The model includes a Drude term and two Lorentz terms, which
approximately capture the contributions by free carriers and
interband transitions. A comparison of the measured $R(\omega)$
and calculated curves at 300 K and 10 K is shown in Fig. 3. In the
inset, the experimental and calculated $R(\omega)$ spectra at 300
K are displayed over broad frequency range. We find that this
simple model can reasonably reproduce the $R(\omega)$ curves. This
is the case particularly for the spectrum at 300 K. At 10 K, the
calculated curve deviates from the experimental spectrum at low
$\omega$. This is understandable, considering that a constant
parameter for scattering rate, 1/$\tau$ in the Drude term, is used
for fitting. In reality, the carrier scattering rate (damping) may
be $\omega$-dependent. The unscreened plasma frequencies,
$\omega_p$ of the Drude term, obtained at different T are shown in
the inset of Fig. 2. The plasma frequency increases steadily from
13,500 \cm to 17,000 \cm with decreasing T, roughly a 20$\%$
increase. The scattering rate keeps roughly constant with
1/$\tau\sim$800-850 \cm. In the analysis, the two Lorentz terms
modelling the interband transitions are centered at 6,500 \cm and
14,500 \cm, respectively. The epsilon infinity obtained is
$\epsilon_\infty$=18. Note that the value of the unscreened plasma
frequency is fairly large, this is mainly due to the very small
value of the effective mass. An estimation of the effective mass
will be given below.

Figure 4 shows the optical conductivity $\sigma_1(\omega$) below
1,000 \cm at different T. A Drude-like conductivity is present at
low $\omega$, and its spectral weight (i.e. the area under the
$\sigma_1(\omega$) curve) increases with decreasing temperature.
This is expected, as the spectral weight of Drude-like
conductivity is linked to effective conducting carrier density,
consistent with the T-dependent evolution of both the reflectance
and energy loss function spectra presented above. The inset shows
the room-T $\sigma_1(\omega$) over a broad energy range. Besides
the Drude component at low frequency, there exist several
interband transition structures. The interband transition near
6,500 \cm appears as a weak shoulder of the prominent peak near
14,500 \cm. Stronger structures also appear at frequencies above
20,000 \cm. All these features can be attributed to interband
transitions from occupied Se 4p bands to unoccupied parts of Ti 3d
bands in different momentum directions in
\textit{\textbf{K}}-space\cite{Zunger,Reshak}.

\begin{figure}[t]
\centerline{\includegraphics[width=2.9in]{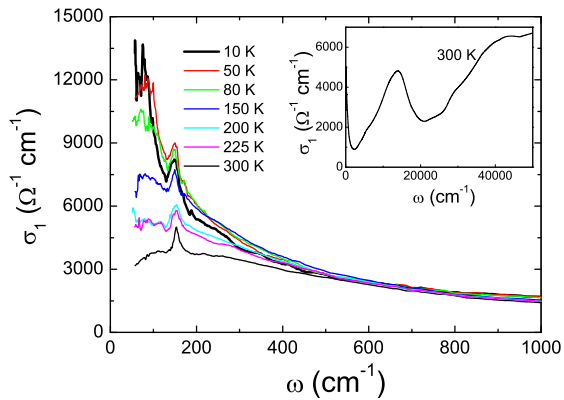}}%
\vspace*{-0.20cm}%
\caption{(Color online) Frequency dependence of the optical
conductivity at different T. The inset shows
the $\sigma_1(\omega$) at 300 K over a broad energy range.}%
\label{fig4}
\end{figure}

From the data presented above, we can see that the most striking
property of this doped CDW material is the blueshift of the plasma
frequency with decreasing T. Such a phenomenon is rarely seen in
metals. To our knowledge, europium hexaboride (EuB$_6$) is the
only example known to date showing a similar blueshift of the
plasma edge\cite{Degiorgi}. However, EuB$_6$ enters into an
ferromagnetically ordered state below T$_c\sim$16 K: the blueshift
of the plasma edge occurs only when the compound is cooled through
the ferromagnetic transition. Moreover, the plasma frequency in
EuB$_6$ shifts to higher energy when a magnetic field is applied.
Thus it is believed that the effect is related to the reduced
scattering by spin ordering, or undressing effect as explained by
Hirsch.\cite{Hirsch} Cu$_{0.07}$TiSe$_2$ is, however, a
superconductor and does not have any magnetic ordering.

It is well known that the plasma frequency is determined by the
charge carrier density and the effective mass. In the case of a
single band, $\omega_p^2=4\pi ne^2/m^*$, the observed blueshift
indicates either a sizeable increase of carrier density or/and a
reduction of the effective mass of the conducting charge carriers.
However, it is unlikely that the carrier density changes with T.
The carrier density is a quantity controlled by the valence
electrons offered by constituent atoms, which should not change
with T if the crystal structure does not undergo a symmetry
breaking transition that could lead to a change of the number of electrons
per unit cell. One exception is a metal or semimetal with very low
Fermi temperature ($T_F$) due to a small band crossing E$_F$
near its top or its bottom; in that case the carrier number could be changed
by thermal activation\cite{Halperin}. However, the number of such
thermally activated carriers can only decrease with decreasing T,
opposite to the trend observed here. Very recently, Hall
effect measurements have been reported for Cu-doped TiSe$_2$
crystals.\cite{Wu} Indeed, it was found that the Hall coefficient
R$_H$ changes with T only for lightly Cu-doped samples, where CDW
transitions exist. For the heavily Cu-doped samples, such as
superconducting Cu$_{0.07}$TiSe$_2$, R$_H$ is T-independent,
which therefore rules out the possibility of a T-dependent carrier
density.

As the carrier density does not change with T, we are left with
the possibility that the effective mass of charge carriers
decreases with decreasing T. From our observed plasma frequencies and the Hall
coefficient results\cite{Wu}, we estimate that a mass reduction, from m$^*$=0.55m$_e$ to 0.35 m$_e$, occurs in  Cu$_{0.07}$TiSe$_2$ on cooling from
300 to 10 K. This behavior is in sharp contrast to many
so-called strongly correlated electron systems that show an
enhanced carrier effective mass at low T due to a many-body
renormalization effect\cite{Degiorgi2,Wang}. In order to get
insight into this issue, it is necessary to examine the evolution
of the band structure with Cu intercalation. The parent compound
1T-TiSe$_2$ is a well known CDW material. Its ground state is
revealed to be either a low carrier density semimetal or a
semiconductor with a very small indirect
gap\cite{Zunger,Bachrach,Traum,Anderson,Pillo,Kidd,Rossnagel,Cui,Li}.
With the intercalation of Cu$^{1+}$ into the structure, extra
electrons are doped into the system. Within a rigid band picture,
this will raise the chemical potential. Indeed, recent ARPES
experiments indicate that Cu-intercalation places the bottom of the
Ti-3d electron band near the L-point further below
E$_F$\cite{Qian,Zhao}. This electron band dominates the electronic
properties, in agreement with the negative Hall and Seebeck
coefficients seen in transport experiments\cite{Morosan,Wu}.

It is noted that the electrons near L-points couple strongly with
lattice.\cite{Holt} Then, a T-dependent electron-phonon coupling
may cause a change of carrier effective mass. As the phonon number
decreases with decreasing T, a reduced scattering of electrons
from phonons (i.e. a mass reduction) is expected. In early studies
on nontransition metals, such as Zn, a temperature shift of the
mass renormalization arising from the electron-phonon interaction
was observed\cite{Sabo} and well explained theoretically by Allen
and Cohen.\cite{Allen} However, this ordinary electron-phonon
interaction has only a very small effect on the mass shift. It can
not explain the substantial change observed here. In fact, the L
phonons are linked to the structural instability in the parent and slightly
Cu-intercalated compounds. If strong phonon softening takes
place at low T, one would expect to see an enhanced incoherent
low-energy quasiparticle, opposite to the ordinary effect. Indeed,
even for the present x=0.07 compound without a static CDW
instability, recent ARPES experiments indicate that the low-T EDC
(energy distribution curve) quasiparticle peak of the L pocket at a
fixed k-point near E$_F$ does not become sharper after removing
thermal effects.\cite{Qian2} Thus, a naive picture for a
reduced electron-phonon interaction is not applicable here.

We propose following mechanism, which involves a T-induced band
effect, for the mass change in this compound. We note that, for a
system with a constant carrier density, the chemical potential is
not fixed but shifts up with decreasing T when the Fermi energy of
the system is not high. Assuming a degenerate electron gas for the
metallic phase, the chemical potential has the following relation
with T, $\mu\simeq\mu_0[1-\pi^2/12(T/T_F)^2]$. For a usual metal,
the Fermi temperature $T_F$ is very high, and the T-induced change
in $\mu$ in the range 0-300 K is negligible. However, for the
present case, the Fermi energy is rather low, about 80-100 meV at
300 K.\cite{Qian2} Then we expect to have an upward shift of
chemical potential by about 5-10$\%$ for T decreasing from 300 K
to 0 K. If the effective mass of electrons is different along the
dispersive band, a change of effective mass at E$_F$ could be
caused by a chemical potential shift. As the L point is linked to
the lattice instability, the electrons at the L point experience
the strongest scattering. Away from the L point, along the
dispersive band, the scattering would be significantly reduced.
Indeed, from the ARPES experiment the EDC width at the L point
(the band minimum) is much broader than other points away from the
L point along the dispersive band.\cite{Qian,Zhao} Thus an upward
shift of chemical potential with decreasing T, driving the Fermi
crossing point further away from the L point, would lead to a
relatively smaller effective mass at E$_F$.

To summarize, we performed optical spectroscopy study on a newly
discovered superconductor Cu$_{0.07}$TiSe$_2$. The study reveals
it to be a low-carrier density metal, consistent with the
development from a semimetal or semiconductor with a small
indirect gap upon doping. Surprisingly, the data reveal a
substantial shift of the plasma edge in reflectance towards high
energy with decreasing T. This phenomenon, rarely seen in metals,
indicates either a sizeable increase of the conducting carrier
concentration or/and a decrease of the effective mass of carriers
with reducing T. Our analysis indicates that the blueshift of the
plasma frequency in optics is mainly caused by the later effect.

We acknowledge very helpful discussions with Z. Q. Wang, L. Yin,
D. L. Feng, Q. H. Wang and Lu Yu. This work is supported by
National Science Foundation of China, the Knowledge Innovation
Project of Chinese Academy of Sciences, and the Ministry of
Science and Technology of China (973 projects). MZH and RJC
acknowledge partial support through NSF(DMR-0213706) and
U.S.DOE/DE-FG-02-05ER46200.

%
%

\end{document}